\documentclass[twocolumn,aps,showpacs]{revtex4-1}
\usepackage[latin9]{inputenc}
\usepackage{float}
\usepackage{textcomp}
\usepackage{amsmath}
\usepackage{amssymb}
\usepackage{graphicx}
\PassOptionsToPackage{normalem}{ulem}
\usepackage{ulem}
\usepackage{subscript}
\makeatletter
\usepackage{color}

\def\NOT(#1,#2){\OneQubitGate(#1,#2){$X$}}

\makeatother

\begin{document}

\title{Experimental characterization of spin 3/2 silicon-vacancy centers
in 6H-SiC}

\author{Harpreet Singh\textsuperscript{1}, Andrei N. Anisimov\textsuperscript{2},
Sergei S. Nagalyuk\textsuperscript{2}, Eugenii N. Mokhov\textsuperscript{2},
Pavel G. Baranov\textsuperscript{2} and Dieter Suter\textsuperscript{1}\\
 \textsuperscript{1}Fakultät Physik, Technische Universität Dortmund,\\
 D-44221 Dortmund, Germany. \textsuperscript{2}Ioffe Institute, St.
Petersburg 194021, Russia.}
\begin{abstract}
Silicon carbide (SiC) hosts many interesting defects that can potentially
serve as qubits for a range of advanced quantum technologies. Some
of them have very interesting properties, making them potentially
useful, e.g. as interfaces between stationary and flying qubits. Here
we present a detailed overview of the relevant properties of the spins
in silicon vacancies of the 6H-SiC polytype. This includes the temperature-dependent
photoluminescence, optically detected magnetic resonance (ODMR) and
the relaxation times of the longitudinal and transverse components
of the spins, during free precession as well as under the influence
of different refocusing schemes.
\end{abstract}
\maketitle

\section{Introduction}

\subsection{Spin centers in silicon carbide}

Silicon carbide (SiC) is known for the diversity of its polytypes
with remarkable and tunable electrical and optical properties as well
as its radiation stability ~\cite{tarasenko-pssb-18,falk-nature-12}.
SiC has a large bandgap with deep defects and is supported by sophisticated
fabrication techniques~\cite{janzen-physb-09,roya-apl-13,song-optexp-11,wang-prap-17}.
Recently, silicon-vacancy (V$_{Si}$) centers in SiC were proposed
as an alternative to Nitrogen vacancy (NV) centers in diamond for
spintronics and quantum technologies~\cite{widmann-nature-14}. The
photoluminescence of the defects in SiC lies in the near infrared,
which allows, e.g., optical detection of spin states~\cite{baranov-jetpl-05,son-sst-99,kraus-nature-13,baranov-prb-11,fuchs-nature-15}.
The color centers in SiC can be grouped into two classes depending
on their spin in the ground state: $S$=1 or $S$=3/2~\cite{sorman-prb-00,tarasenko-pssb-18,mizuochi-prb-02,bardeleben-prb-00,wagner-prb-02,son-sst-99,carlos-prb-06,Orlinski-prb-03}.
The divacancies (${\rm {V_{Si}-V_{C}}}$) are formed by adjacent pairs
of Si and C vacancies, have spin $S$=1 and are known as P6 and P7
in the literature~\cite{son-prl-06,baranov-jetpl-05,son-sst-99,christle-nature-14,lingner-prb-01}.

\begin{figure}
\includegraphics{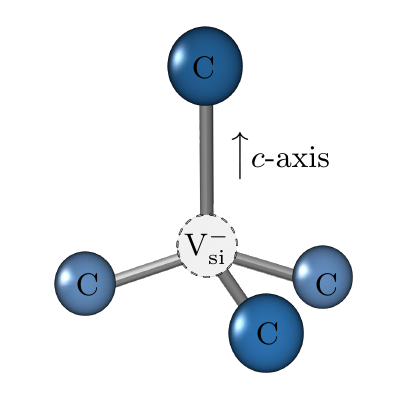}

\caption{Structure of a negatively charged spin 3/2 silicon-vacancy (V$_{Si}^{-}$).
The blue spheres represent the carbon atoms. The white circle with
the dashed boundary represents the V$_{Si}^{-}$.}

\label{c3vsatructure}
\end{figure}

The four dangling sp$^{3}$ orbitals at the V$_{Si}$ site contribute
four electrons. If V$_{Si}$ captures an additional electron, it becomes
a negatively charged silicon vacancy (V$_{Si}^{-}$)~\cite{baranov2013,soykal-prb-16,anisimov-aipa-2018,wagner-prb-02}
whose spin state was shown to be $S$ = 3/2~ by a radiofrequency-optical
double resonance technique \cite{riedel-prl-12,soykal-prb-16}. The
site symmetry of V$_{Si}^{-}$ is C$_{3v}$, as shown in Fig.$\,$\ref{c3vsatructure}.
Several separately addressable V$_{Si}^{-}$ have been identified
in the same crystal for each of the main SiC polytypes: hexagonal
4H-SiC and 6H-SiC and rhombic 15R-SiC. The 4H-SiC polytype, e.g.,
hosts one hexagonal ($h$) and one cubic ($k$) lattice site and in
6H-SiC there are one hexagonal and two cubic sites ($h$, $k_{1}$
and $k_{2}$). V$_{Si}^{-}$ at a hexagonal site $h$ of 4H- and 6H-SiC
is called a V$_{2}$ type vacancy, at a cubic site $k$ of 4H-SiC
it is called V$_{1}$ and in 6H-SiC V$_{1}$ and V$_{3}$ are located
at sites ${k_{1}}$ and ${k_{2}}$~\cite{sorman-prb-00}.

The spins of V$_{Si}^{-}$ in SiC are highly controllable and can
be manipulated with the techniques that have been developed, e.g.,
for working with diamond NV qubits. This was demonstrated with ensembles~\cite{koehl-nature-11}
as well as with single centers~\cite{widmann-nature-14}. Optically
induced spin polarisation of the ground state at room temperature
has been demonstrated using electron spin resonance~\cite{soltamov-prl-12}.
This spin polarisation can be used to implement solid-state masers
and extraordinarily sensitive radio-frequency amplifiers~\cite{kraus-nature-13}
or magnetic field sensors with dc field sensitivities $>$
${\rm {100~nT/\sqrt{Hz}}}$~\cite{simin-prx-16}. In some V$_{Si}^{-}$
the zero-field splitting (ZFS) is nearly temperature independent,
making these centers very attractive for vector magnetometry. Contrarily,
the zero-field splitting of the centers V$_{2}$ centers in 4H-SiC
in the excited state exhibits a large thermal shift, which makes them
useful for thermometry applications~\cite{anisimov-sr-2018}. All
four ground state spin levels of V$_{Si}^{-}$ have been used to demonstrate
absolute dc magnetometry, which is immune to thermal noise and strain
inhomogeneity~\cite{soltamov-naturecom-19}.

\subsection{Preserving spin coherence}

An important precondition for the implementation of quantum technologies,
including quantum information processing is that the phase of superposition
states can be preserved for times significantly longer than the duration
of the computational task~\cite{divincenzo,Stolze:2008xy}. Achieving
this goal generally requires an array of measures~\cite{RevModPhys.88.041001},
including passive schemes like decoherence-free subspaces~\cite{3045}
or active schemes like spin-echoes~\cite{2028}. In divacancy spin
ensembles in 4H-SiC, Hahn-echo times longer than 1 ms were observed
at 20 K~\cite{christle-nature-14}.

In most cases, the interactions with the environment that cause the
dephasing are not static and a single refocusing pulse is not sufficient
to completely suppress the dephasing. For those cases, multiple echo
sequence were developed by Carr and Purcell~\cite{carr-pr-1954},
consisting of equidistant $\pi$ pulses. This basic experiment was
improved to make it more robust with respect to experimental imperfections~\cite{meiboom:688,souza-prl-11,souza-ptrsa-12}.
These so-called dynamical decoupling (DD) techniques have been applied
to extend the coherence times of different types of qubits, including,
e.g., rare-earth ions~\cite{Zhong:2015fk} and the spin of the NV
center in diamonds~\cite{Shim-epl-2012}. In the V$_{2}$ vacancy
of the 4H-SiC polytype, the efficiency of the spin-echo experiment
depends strongly on the magnetic field~\cite{carter-prb-15}. Combining
Carr-Purcell-Meiboom-Gill (CPMG) decoupling with a static magnetic
field can extend the spin coherence time of the V$_{2}$ center in
4H-SiC to more than 20 ms~\cite{simin-prb-17}.

\subsection{Outline of this paper}

In this work, we focus on the 6H-SiC polytype whose spin properties
have not yet been studied in detail. Sec.~\ref{system} gives details
of the sample preparation, photoluminescence measurement, and the
optical pumping scheme. Sec.~\ref{sec:expresult} describes the experimental
setup for continuous-wave as well as pulsed optically detected magnetic
resonance (ODMR) measurements. Sec.~\ref{sec:relaxmeasure} describes
the results of the spin-lattice and spin-spin relaxation measurements.
Sec.~\ref{conc} contains the discussion and concluding remarks.

\section{System}

\label{system} 

\subsection{Sample}

The experiments were performed on a sample that was isotopically enriched
in $^{28}$Si and $^{13}$C. The Si enrichment was performed using
gas-centrifuge technology, resulting in an isotopic purity of 99.999\%,
which was available in the form of small $^{28}$Si pieces (1-3 mm).
The \textsuperscript{13}C source was carbon powder enriched to 15\%
in \textsuperscript{13}C. The SiC crystal was grown at a temperature
of 2300\textdegree -2400\textdegree C on a (0001) Si face in an Argon
atmosphere at a growth rate of $\approx$ 100$\mu$m/h. After growing
the SiC crystal, machining and cutting of the wafers were carried
out. The isotope composition was measured by Secondary Ion Mass Spectroscopy(SIMS).
The concentrations of $^{28}$Si, $^{29}$Si, and $^{30}$Si are 99.918
\%, 0.076\%, and 0.006 \%, respectively. The concentrations of $^{12}$C
and $^{13}$C are 95.3\% and 4.7\%, respectively. To create V$_{Si}^{-}$
centers, the crystal was irradiated with electrons with a dose of
10$^{18}$cm$^{-2}$ and an energy of 2 MeV at room temperature.

\label{sec_PL} 
\begin{figure}
\includegraphics{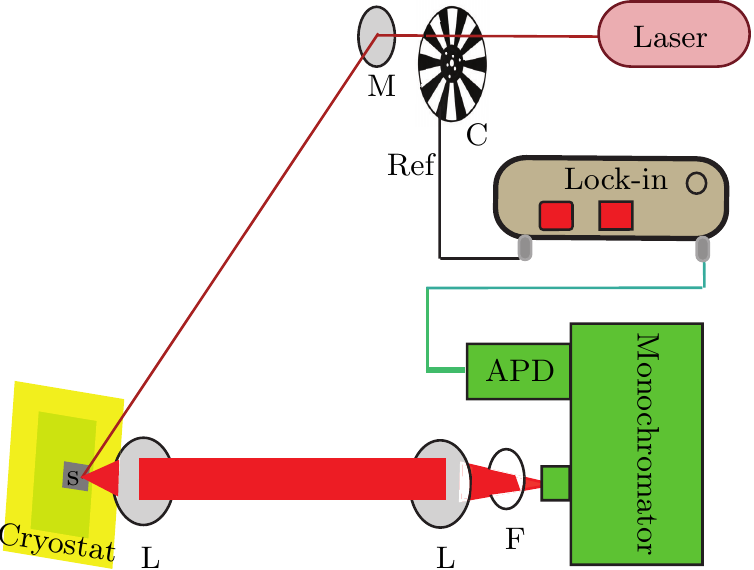} \caption{Experimental setup for photoluminescence measurements. The red line
represents the laser beam, C marks the mechanical chopper. The gray
rectangle labeled S is the SiC sample. Ellipsoids labeled M, L and
F represent reflecting mirrors, convex lenses, and a long-pass filter,
respectively. The rectangle labeled APD represents the avalanche photodiode
module.}
\label{gaas_setup_fig} 
\end{figure}


\begin{figure}
\includegraphics[width=0.99\columnwidth]{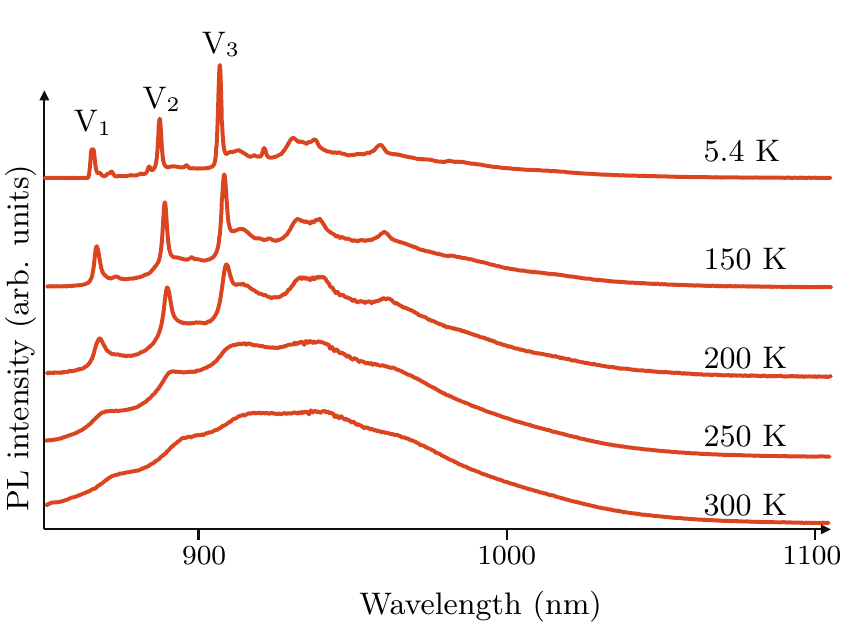} \caption{PL spectra measured at different temperatures. The sample is excited
with a 790 nm laser. Peaks labeled V$_{1}$, V$_{2}$ and V$_{3}$
correspond to the zero-phonon lines of V$_{Si}^{-}$ at the lattice
sites \textit{k$_{1}$}, \textit{h }and \textit{k}$_{2}$ respectively.}
\label{PL_plots} 
\end{figure}

\subsection{Photoluminescence}

To measure the Photoluminescence (PL) of the sample, we used the setup
shown schematically in Fig.~\ref{gaas_setup_fig}. A Ti:sapphire
laser provided the optical excitation. The PL was collected with the
help of two convex lenses of focal length 20 cm. The PL was collected
in the direction of the $c$-axis and passed through a long-pass 850
nm filter (F; Thorlabs) to a monochromator (Spex 1704). An avalanche
photodiode (APD) module with a frequency bandwidth from DC to 100
kHz (C5460-1 series from Hamamatsu) was attached to the monochromator
for detecting the PL. The voltage output of this APD was measured
with the lock-in amplifier (SRS model SR830). The laser beam was modulated
with a chopper, whose sync signal served as reference for the lock-in
amplifier.

Fig.~\ref{PL_plots} shows some PL spectra recorded at different
temperatures with the excitation laser set to 790 nm. The sample was
cooled down using a Helium cryostat, and the PL spectra recorded at
$\approx$ 5.4 K, 150 K, 200 K, 250 K and 300 K are shown. The zero
phonon lines (ZPL) of the negatively charged vacancies are visible
at the expected wavelengths of 865 nm (V$_{1}$), 887 nm (V$_{2}$)
and 908 nm (V$_{3}$)~\cite{sorman-prb-00,wagner-prb-00}.

\subsection{Energy levels and optical pumping}

\label{sec:ODMR}

\noindent 
\begin{figure}[H]
\centering \includegraphics{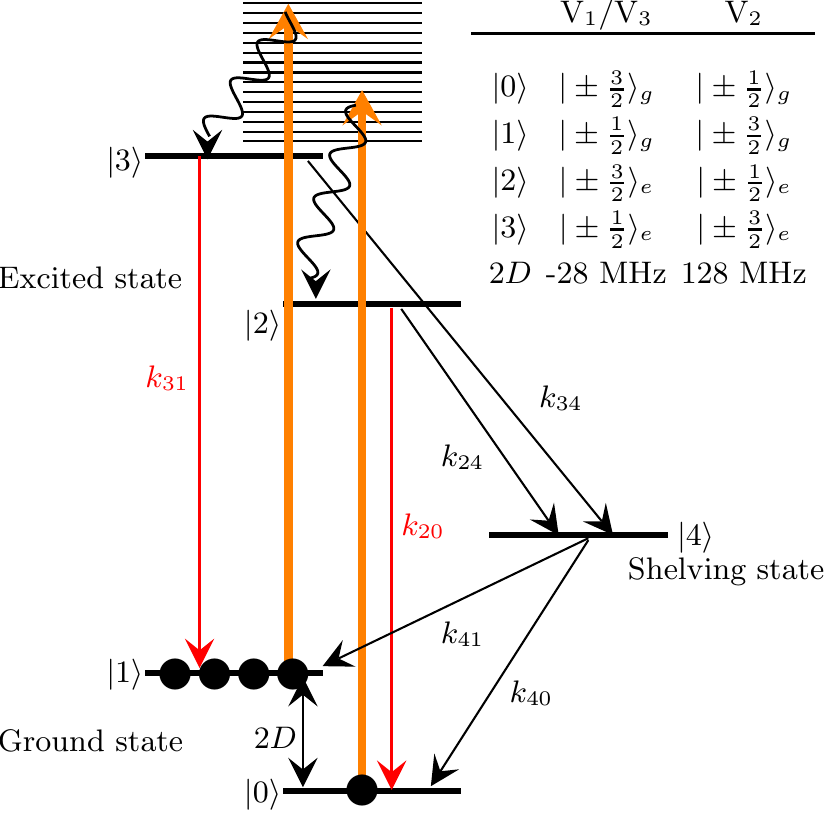} \caption{Energy-level diagram of the 6H-SiC V$_{Si}^{-}$ showing the ground,
excited and shelving states. Radiative transitions are marked by red
arrows. The laser beam excitation is shown with orange arrows. Spin
dependent non-radiative transitions generating the ground-state spin
polarization are shown as black arrows. States $\vert0\rangle$ and
$\vert1\rangle$ represent the degenerate$\vert\pm\frac{3}{2}\rangle$(
$\vert\pm\frac{1}{2}\rangle$ ) and $\vert\pm\frac{1}{2}\rangle$
($\vert\pm\frac{3}{2}\rangle$ ) ground states of the V$_{1}$/V$_{3}$
(V$_{2}$) type $V_{Si}^{-}$. The states $\vert2\rangle$ and $\vert3\rangle$
represent the doubly degenerate excited states and $\vert4\rangle$
the shelving states.}
\label{energy_levels} 
\end{figure}

The negatively charged defects in 6H-SiC have spin $S$ = 3/2~\cite{riedel-prl-12,soykal-prb-16}.
Fig.~\ref{energy_levels} shows the energy-level diagram of the 6H-SiC
V$_{Si}^{-}$ in the absence of an external magnetic field. The states
$\vert0\rangle$ and $\vert1\rangle$ in the electronic ground state
correspond to  the $S=3/2$ $m_{S}=\pm\frac{3}{2}$ and $\pm\frac{1}{2}$
spin states. In the absence of a magnetic field, they form two degenerate
doublets, which are split by the zero-field interaction. In the case
of the V$_{1}$/V$_{3}$ vacancy, the $m_{S}=$ $\pm\frac{3}{2}$
are lower in energy, i.e. they correspond to state $|0\rangle$, in
the $V_{2}$ vacancy, the $\pm\frac{1}{2}$ states are the lowest
energy states~\cite{kraus-nature-13,biktagirov-prb-18}. The states
$\vert2\rangle$ and $\vert3\rangle$ represent the $S=3/2$, $m_{S}=$
$\pm\frac{3}{2}$($\pm\frac{1}{2}$ ) and $\pm\frac{1}{2}$ ($\pm\frac{3}{2}$)
spin substates of the electronically excited states of V$_{1}$/V$_{3}$
(V$_{2}$)~\cite{baranov-prb-11,fuchs-nature-15}. The shelving state
$|4\rangle$ is an $S=1/2$ state, which is important for the optical
pumping process ~\cite{baranov-prb-11}.

\begin{figure}[H]
\includegraphics{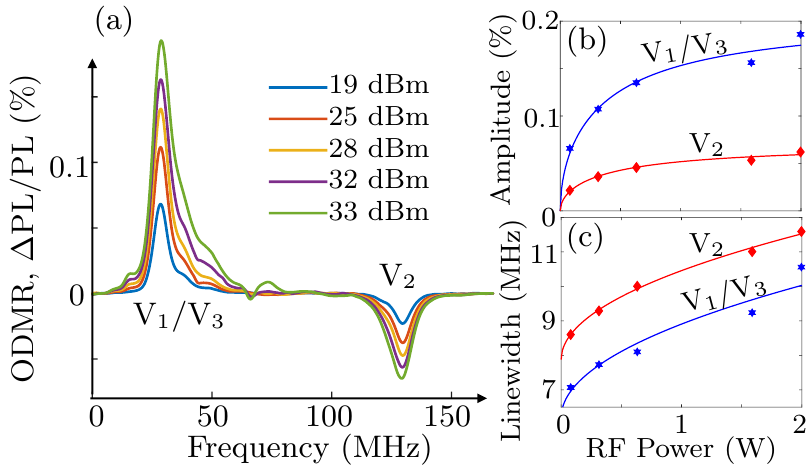} \caption{(a) ODMR signal vs. frequency recorded with different RF powers in
zero magnetic field. The horizontal axis is the frequency in MHz and
the vertical-axis the relative change of the PL, recorded by the lock-in
amplifier. (b) ODMR signal vs. RF power (c) Linewidth vs. RF power.}
\label{odmr_plot}
\end{figure}

The spin Hamiltonian of the $S=3/2$ states is 
\begin{equation}
\mathcal{H}=D(S_{z}^{2}-5/4)+g\mu_{B}\vec{B}\cdot\text{\ensuremath{\vec{S}}},\label{hamiltonian}
\end{equation}
where the zero field splitting in the electronic ground state is $2D=-28$
MHz for V$_{1}$/V$_{3}$ and 128 MHz for V$_{2}$~\cite{biktagirov-prb-18},
$g=2.0$ is the electron $g$-factor, $\mu_{B}$ is the Bohr magneton,
$\vec{B}$ is the external magnetic field, $S$ is the vector of the
electron spin operators. We use a coordinate system whose $z$-axis
is oriented along the C$_{3}$ symmetry axis, which is also the $c$-axis.
In the absence of optical pumping, when the spin system is in thermal
equilibrium at room temperature, all four ground states are almost
equally populated. When the system is irradiated with a laser, the
populations are re-distributed, as shown schematically in Fig.~\ref{energy_levels}.
When the laser is turned on, it excites transitions from the ground
states $\vert0\rangle$ and $\vert1\rangle$ to the excited states
$\vert2\rangle$ and $\vert3\rangle$. From the excited states $\vert2\rangle$
and $\vert3\rangle$ most of the population falls back to the $\vert0\rangle$
and $\vert1\rangle$ states by spontaneous emission with a rate $k_{20}$
and $k_{31}$. However, the system can also undergo intersystem-crossing
(ISC) to the shelving states $|4\rangle$ with the rates $k_{24}$
and $k_{34}$~\cite{baranov-prb-11}. From there the system returns
to the ground state, with a bias for the state $\vert1\rangle$ over
state $\vert0\rangle$ with the rates $k_{40}$ and $k_{41}$~\cite{riedel-prl-12,biktagirov-prb-18,soltamov-naturecom-19}.
The exact ISC rates from and to the shelving state are not yet known
precisely but, by considering the recorded ODMR data shown in Fig.
\ref{odmr_plot}, $k$$_{34}$ \textgreater{} $k$$_{24}$ for V$_{1}$/V$_{3}$
and $k_{24}$ \textgreater{} $k_{34}$ for V$_{2}$.

\noindent 

\section{Optically detected magnetic resonance}

\label{sec:expresult}

\begin{figure}
\includegraphics[scale=0.85]{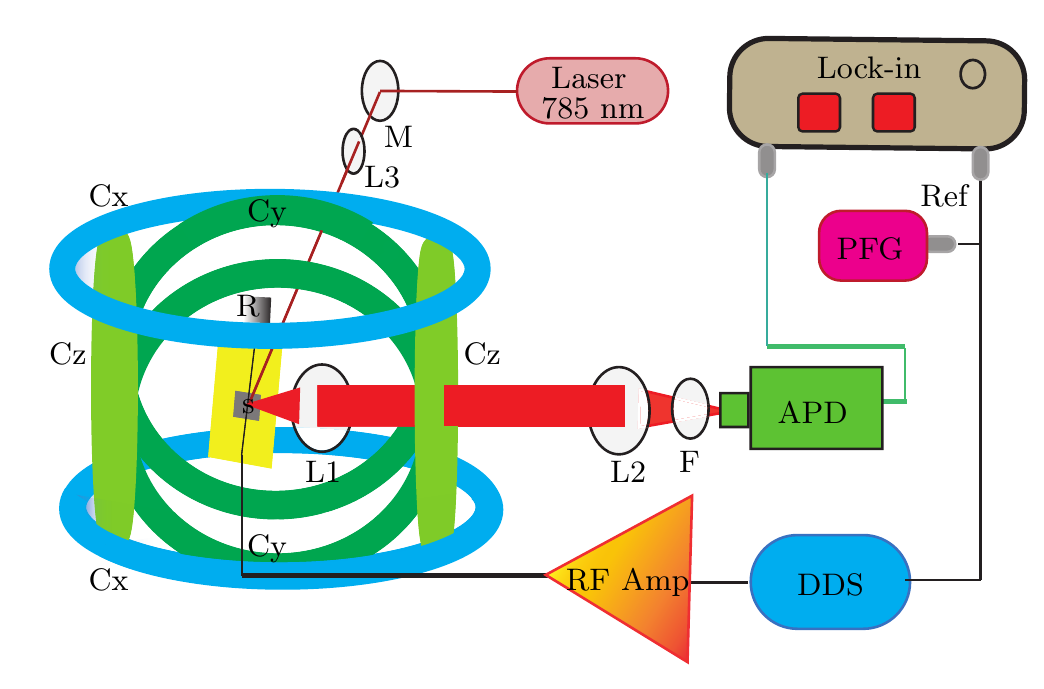} \caption{Experimental setup for measuring the ODMR of silicon vacancies. The
red line represents the path of the laser beam. Ellipsoids labeled
M, Land F represent mirrors, convex lenses, and long-pass filter receptively.
The gray rectangle labeled with S is the SiC sample. The RF is applied
using straight a 50 $\mu$m diameter copper wire placed over the sample,
in series with a 50-Ohm resistor which is represented by a rectangle
labeled with R. The tree orthogonal ring-pairs Cx, Cy and Cz represent
Helmholtz coils. They allows us to apply magnetic fields in an arbitrary
direction. Rounded rectangles labeled PFG, APD and DDS represent a
programmable function generator, an avalanche photodetector module
and a direct digital synthesizer, respectively.}
\label{odmr_setup}
\end{figure}

\subsection{Continuous-wave ODMR}

To determine the ground state spin Hamiltonian of the V$_{Si}^{-}$
in the sample, we used the continuous-wave (cw) ODMR technique with
the setup shown in Fig.~\ref{odmr_setup}. Our light source was a
785 nm laser diode with a maximum power of 400 mW, which was driven
by a Thorlabs laser diode controller (LDC202C series) with a thermoelectric
temperature controller (TED 200C). We used three orthogonal Helmholtz
coil-pairs for applying the static magnetic field in an arbitrary
direction. A highly stable linear current source (Servowatt, three-channel
DCP-390/30) delivers currents up to 15A to the coils. The currents
were controlled individually by an analog control voltage. The radio-frequency
(RF) signal was generated with a direct digital synthesizer (DDS)
AD9915 from Analog Devices which generates signals up to 1 GHz. Its
output was amplified using an RF amplifier (Mini- Circuit LZY-1, 50W
amplifier with a frequency range from 20 MHz to 512 MHz) and sent
to a 50 $\text{ \ensuremath{\mu}m}$ wire terminated with a 50 $\Omega$
resistor. A programmable function generator (PFG, Hameg model HM8130-2)
was used to modulate the amplitude of the RF field. A laser beam was
focused on the sample using a convex lens (L3) of focal length 20
cm. The PL from the sample was collected with a pair of lenses (L1
and L2 with focal lengths 5 cm and 15 cm respectively), sent through
a 850 nm long-pass filter to suppress stray light from the laser and
to an avalanche photodiode (APD) module with a frequency range from
DC to 10 MHz (C12703 series from Hamamatsu). The APD signal was demodulated
with a lock-in amplifier (SRS model SR830 DSP) whose reference signal
was derived from the PFG modulating the RF.

\noindent 
\begin{figure}[h]
\includegraphics[width=0.99\columnwidth]{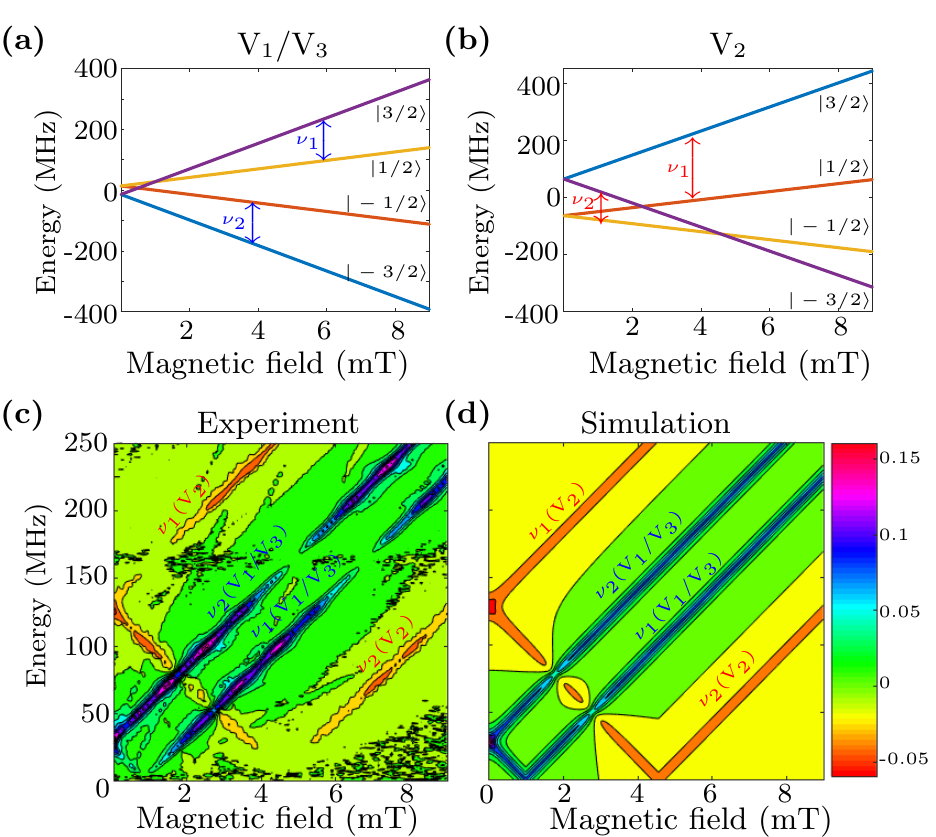} \caption{Energy levels of (a) theV$_{1}$/V$_{3}$ vacancy and (b) V$_{2}$
vacancy in a magnetic field $B$ $\parallel$ $c$-axis. (c) Experimental
ODMR and (d) Simulated ODMR showing resonances from V$_{1}$/V$_{3}$
and V$_{2}$ for a range of magnetic fields $B$ $\parallel$ $c$-axis.
The color scale in (c) and (d) is in units of ${\rm {\Delta PL/PL\%}}$.}
\label{odmr_plots}
\end{figure}
 Fig.~\ref{odmr_plot}(a) shows the ODMR signal recorded in the absence
of a magnetic field by sweeping the frequency of the RF at different
RF powers, using the setup shown in Fig.~\ref{odmr_setup}. Two peaks
with different signs are observed: a positive one (i.e. increase of
PL at the application of RF) at 28 MHz and a negative one at 128 MHz.
In a previous work~\cite{sorman-prb-00}, it was shown that the peak
at 128 MHz corresponds to the V$_{Si}^{-}$ at lattice sites $h$
(${\rm {V_{2}}}$ type)~and the peak at 28 MHz corresponds to V$_{Si}^{-}$
at two quasi-cubic sites $k_{1}$ and $k_{2}$ (V$_{1}$ and V$_{3}$
type) which have the same $D$ value~\cite{sorman-prb-00}. Recently
it has been shown the peak at 28 MHz corresponds to V$_{Si}^{-}$
at quasi-cubic site $k_{1}$ and the peak at 128 MHz corresponds to
the V$_{Si}^{-}$ at the quasi-cubic site $k_{2}$ (V$_{3}$ type)~\cite{biktagirov-prb-18}.
Since the ODMR peaks assignment is still debatable, we will follow
the ODMR peaks assignment used in earlier work~\cite{sorman-prb-00}.

The overall ODMR contrast is relatively small, but comparable to the
values from similar systems (see, e.g.,\cite{soltamov-naturecom-19,simin-prb-17,kraus-nature-13,anisimov-sr-2018}).
While not all the relevant parameters are known, one reason for the
relatively small contrast is that the PL from the different types
of vacancies can not be separated at room temperature, as shown in
Fig. \ref{PL_plots}. The measured PL therefore includes baseline
contributions from other centers that do not depend on the magnetic
resonance.

\noindent The variation of amplitude and linewidth with RF power is
shown in Fig.~\ref{odmr_plot} (b) and (c), respectively. The amplitude
data were fitted with the function

\[
S(P)=S_{max}(P/(P_{0}+P)),
\]
where $S(P)$ is the signal amplitude and $P$ the RF power. $S_{max}$
and $P_{0}$ are the fitting parameters and the resulting values were
0.2087$\%$ and 0.8573 $W$ (0.07112 $\%$ and 0.8834 $W$) for V$_{1}$/
V$_{3}$ (V$_{2}$) respectively. The linewidth data were fitted to
the function

\[
LW(P)=LW_{0}+a\sqrt{P},
\]
where $LW(P)$ is the linewidth. The resulting values for the fitting
parameters $LW_{0}$ and $a$ were 6.193 MHz and 2.713 MHz W$^{-1/2}$
(7.877 MHz and 2.579 MHz W$^{-1/2}$) for V$_{1}$/V$_{3}$ (V$_{2}$)
respectively. At the maximum RF power that we could apply, 33 dBm,
the ODMR signal ${\rm {\Delta PL/PL}}$ reached an amplitude of 0.19
\% (-0.06 \%) for V$_{1}$/V$_{3}$ (V$_{2}$) and the linewidth of
V$_{1}$/V$_{3}$ (V$_{2}$) was 10.24 MHz (11.70 MHz).

Figures~\ref{odmr_plots} (a) and (b) shows the energy levels of
V$_{1}$/V$_{3}$ and V$_{2}$ as a function of the magnetic field
$B$ applied $\parallel$ $c$-axis, calculated from the Hamiltonian
given in Eq.~\eqref{hamiltonian}. Arrows labeled with $\nu{}_{1}$
and $\nu{}_{2}$ represent the transition from $\vert3/2\rangle\longleftrightarrow\vert1/2\rangle$
and $\vert-3/2\rangle\longleftrightarrow\vert-1/2\rangle$ respectively.
Fig.~\ref{odmr_plots} (c) summarises the ODMR spectra for a range
of magnetic fields from 0 to 9 mT, applied parallel to the $c$-axis,
using an RF power of 32 dBm (1.6 W). In this plot the magnetic field
strength $B$ is plotted along the horizontal axis and the vertical
axis corresponds to the RF frequency. The relative change of the PL
is color-coded as shown by the color bar to the right of the plot.
For frequencies \textless{} 20 MHz, the RF power generated by the
available amplifier drops significantly, which leads to the very small
ODMR signal in this range. Also, at a frequency of $\sim$165 MHz,
we observed very small signals , which appears to be do to a standing
wave in our RF system, which was not impedance-matched to the 50 $\Omega$
amplifier. The experimental data compare well to the superposition
of the signals from the two types of Si vacancies, which is shown
in Fig.~\ref{odmr_plots}~(d). For this simulation, the transition
frequencies are obtained from the Hamiltonian of Eq.~\eqref{hamiltonian},
while amplitudes and linewidths are taken from the experimental data.

\begin{figure}[h]
\includegraphics[scale=0.8]{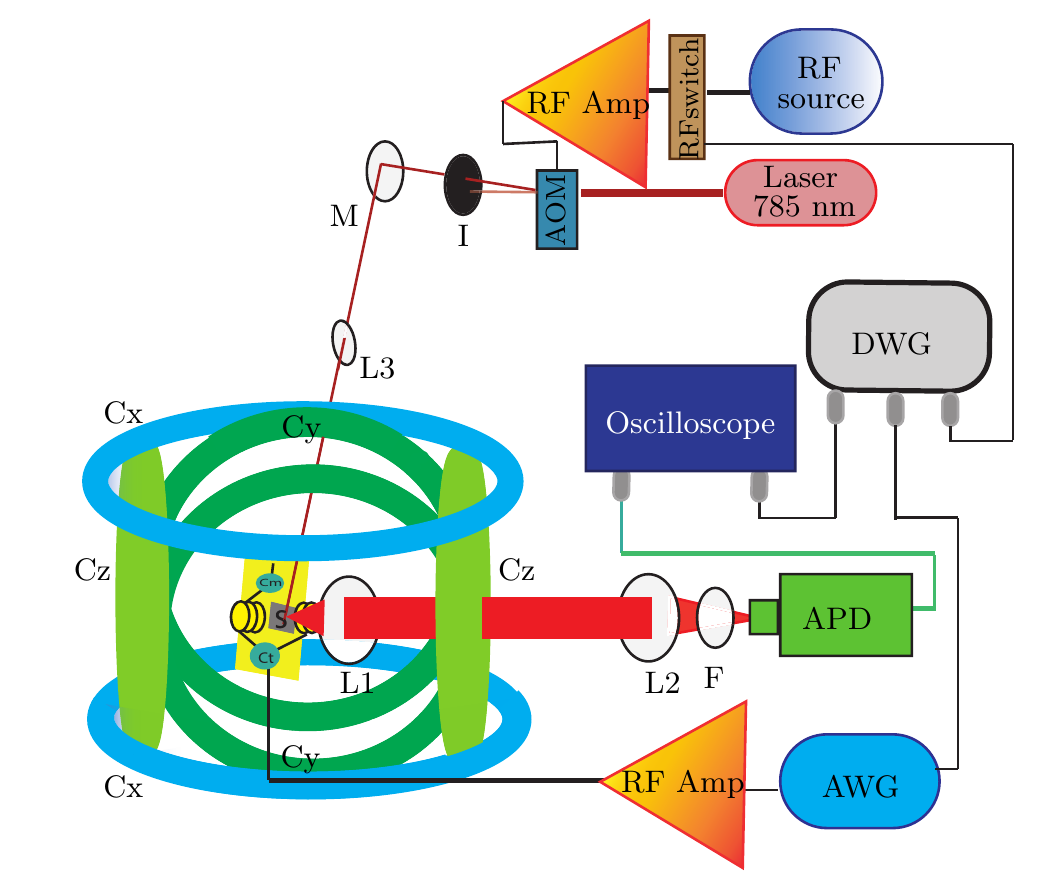} \caption{Experimental setup used for measuring the relaxation rates. The acousto-optical
modulator (AOM) generates the laser pulses. The red line represents
the path of the laser beam. Ellipsoids labeled with M, L and F represent
reflecting mirrors, convex lenses, and long-pass filter receptively.
DWG is a digital word generator (TTL pulse generator). The gray rectangle
labeled with S is the SiC sample. AWG is an arbitrary waveform generator.
The RF is applied to the sample by a resonant LC circuit.}
\label{pulse_setup_fig}
\end{figure}

\subsection{Pulsed ODMR}

\noindent The uncontrolled interaction with a noisy environment has
two effects on a system that has been excited from its thermal equilibrium
state: it causes dephasing and a return of the system to the thermal
equilibrium state. If the system is used in quantum technology applications
~\cite{nielsen-book-02,Stolze:2008xy}, both effects are unwanted
since long coherence times are an essential requirement for technologies
like quantum computing, quantum metrology and quantum memories. To
assess the suitability of for these applications, a precise characterisation
of the decoherence processes is therefore essential. For this purpose,
we use the time-resolved ODMR technique~\cite{Depinna_1982,langof-jpcb-02}.

\noindent For the time-resolved measurements, we modified the setup
of Fig.~\ref{odmr_setup} to generate laser- and RF pulses of variable
duration. Fig.~\ref{pulse_setup_fig} shows the modified setup. For
generating the laser pulses, we used an acousto-optical modulator
(AOM; NEC model OD8813A). The center frequency of the AOM was 100
MHz, and the RF power was 1.2 W. The RF control signal was generated
by a programmable 1 GHz synthesizer HM8133-2 and the RF pulses were
generated by an RF switch (Mini-Circuits ZASWA-2-50DR+, DC-5 GHz).
The TTL pulses that control the timing were generated by a SpinCore
PulseBlaster ESR-PRO PCI card. The RF pulses that drive the spins
were generated by an AWG (WavePond - Chase Scientific, DAx14000).
The RF pulses were amplified up to 50 W and applied to the SiC sample
through a tuned circuit for minimizing reflections. The signal from
the APD detector was recorded with a USB card (PicoScope 2000 series)
attached to a computer.

\noindent In all experiments described below, a laser pulse of 75
mW power and 300 $\mu$s duration initializes the SiC vacancies by
populating the state $\vert1\rangle$ more than the state $\vert0\rangle$.
At this laser intensity, the time constant for the transfer of population
to spin state $\vert1\rangle$ is 28 $\mu$s (11 $\mu$s) for V$_{2}$
(V$_{1}$/V$_{3}$). After the polarisation of the spin system, a
sequence of RF pulses was applied to the system, as discussed in detail
below. To read out the final state of the spin system, we applied
a second laser pulse of duration 4 $\mu$s during which we collect
the PL as described in Sec.~\ref{sec:expresult}A. In the time-resolved
experiment, we averaged the signal 500 times and subtracted it from
a 500 times averaged signal of a reference pulse sequence to remove
unwanted background signals. This process was repeated 5 times and
again the average was taken .

\begin{figure}[h]
\includegraphics{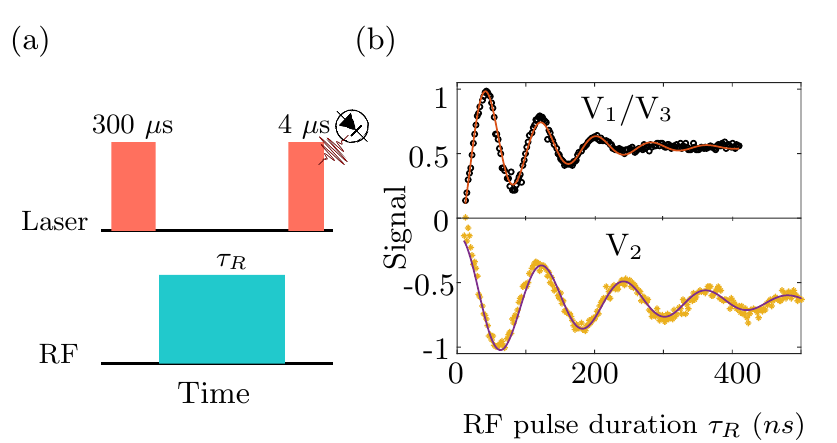} \caption{(a) Pulse sequence for measuring Rabi oscillations. The red and blue
rectangles represent the laser and RF pulses and the pulse duration
is written above the pulse. (b) Experimental Rabi oscillations for
V\protect\textsubscript{1}/ V$_{3}$ and V\protect\textsubscript{2}.
The $y$-axis represents the normalised change of the PL signal and
the $x$-axis the RF pulse duration \textit{$\tau_{R}$}.}
\label{Rabi_all}
\end{figure}

To calibrate the strength of the RF field for the pulsed excitation,
we performed a measurement of Rabi oscillations for the transition
between the $\vert0\rangle$ and $\vert1\rangle$ of $V_{Si}^{-}$,
using the pulse sequence shown in Fig.~\ref{Rabi_all} (a). After
this initializing laser pulse, a 16 W RF pulse of variable duration
$\tau_{R}$ was applied. Here the reference signal was obtained from
an experiment without an RF pulse. Fig.~\ref{Rabi_all} (b), shows
the resulting experimental data for the V\uline{}\textsubscript{1}/
V$_{3}$ and V\uline{}\textsubscript{2} type vacancies. The experimental
data were fitted to the function

\begin{equation}
S_{RF}(\tau_{R})-S_{0}(\tau_{R})=A+B\text{ }cos(2\pi\nu_{R}\tau_{R}-\phi)e^{-\tau_{R}/T_{2}^{\text{*}}},\label{eq_rabi-1}
\end{equation}
where \textit{S}\textsubscript{\textit{RF}}\textit{($\tau_{R}$)
}is the signal measured with an the RF pulse of duration\textit{ $\tau_{R}$
}and \textit{S}\textsubscript{0}\textit{($\tau_{R}$)} the reference
signal without RF\textit{ }pulse\textit{.} For V\uline{}\textsubscript{1}/V$_{3}$,
we obtained the fit parameters $A$ = $0.54$, $B$ = $-0.66$ , $\phi$=
$0.06\pi$, $\nu_{R}$ = $12.44$ MHz, $T_{2}^{\star}$ \textsuperscript{}=
$99.29$ ns and for V\uline{}\textsubscript{2} $A$ = $0.65$,
$B$ = $0.53$, $\phi$= $-0.08\pi$ $\nu_{R}$ = $8.36$ MHz, $T_{2}^{*}$\textsuperscript{}=
$204.81$ ns.

\section{Relaxation measurements}

\label{sec:relaxmeasure}

\subsection{Population relaxation}

\begin{figure}[h]
\includegraphics{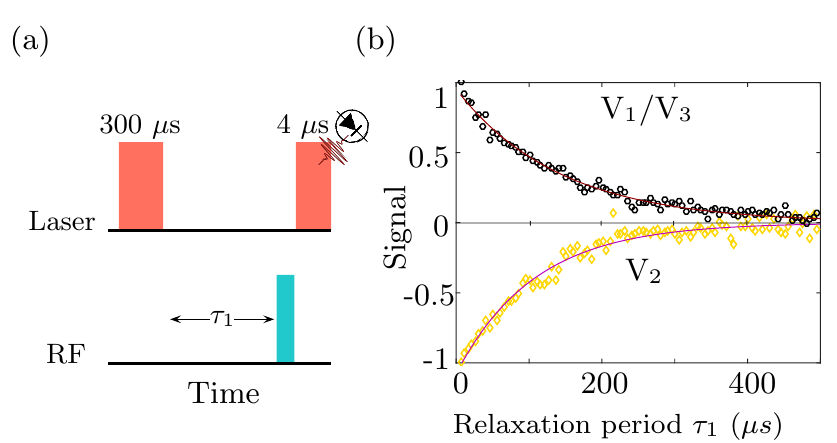} 

\caption{(a) Pulse sequence used to measure the $T_{1}$ relaxation. The red
and blues rectangles represent the laser - and RF pulses. The length
of the pulse is written above the pulse. (b) Resulting signal (normalised)
as a function of the delay $\tau_{1}$, measured at room temperature.}
\label{t1_all}
\end{figure}

Spin-lattice relaxation is the process by which the populations of
the spin system relax to the thermal equilibrium state. It is also
known as longitudinal relaxation and contributes to the decoherence
process. The pulse sequence used to measure the longitudinal relaxation
is shown in Fig.\ref{t1_all} (a). After the initializing laser pulse
the system was allowed to relax for a time $\tau_{1}$ and then the
measuring laser pulse was applied to record the remaining population
difference. The result of this experiment was subtracted from a similar
experiment where the populations of the levels $\vert0\rangle$ and$\vert1\rangle$
were inverted by a $\pi$ pulse applied to the transition between
them. Fig.\ref{t1_all} (b) shows the resulting signals for both vacancy
spins as a function of the delay $\tau_{1}$. The experimental signal
was fitted to the function
\[
S_{\pi}(\tau_{1})-S_{0}(\tau_{1})=A~e^{-\tau_{1}/T_{1}}
\]
where $S_{\pi}(\tau_{1})$ and $S_{0}(\tau_{1})$ are the average
signal measured during the reference and the main pulse sequence respectively
for different delays $\tau_{1}$. From the fit, we obtained the $T_{1}$
relaxation times 142.1$\pm$3.6 $\mu$s and 107$\pm$6.6 $\mu$s for
the V$_{1}$/V$_{3}$ and V$_{2}$ sites, respectively.

\subsection{Free Induction Decay}

\begin{figure}[h]
\includegraphics{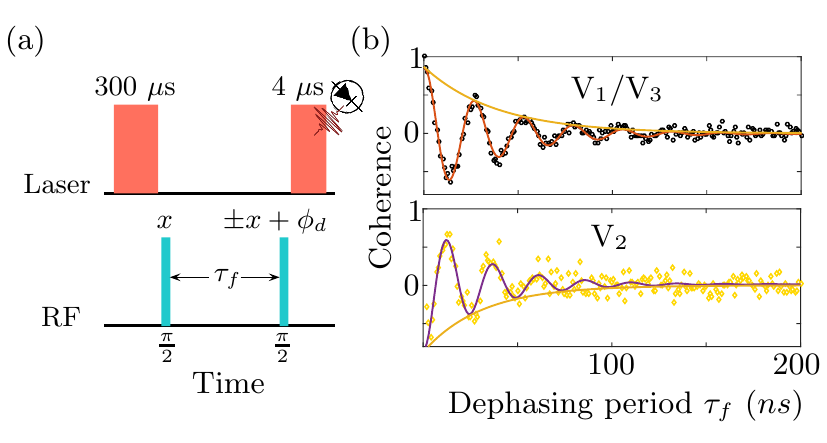}

\caption{(a) Pulse sequences for measuring the free-induction decay. The red
and blues rectangles represent the laser pulses and RF pulses respectively.
The pulse duration is written above the pulse. (b) FID signals measured
for V$_{1}$/V$_{3}$ and V$_{2}$.}

\label{fid_all}
\end{figure}

Another important process is the decay of coherence, which can be
observed in a free induction decay (FID) measurement. While the free
precession of spin coherence can be observed directly in conventional
magnetic resonance, here we used the Ramsey scheme~\cite{ramsey-pr-50}
where a $\pi/2$ RF pulse converted the coherence into a population
difference, which was then read out during the final laser pulse.
Figure~\ref{fid_all} shows the experimental scheme: After the initialisation
by the first laser pulse, the first RF pulse generated the coherence,
which was then allowed to precess for a time $\tau_{f}$ before it
was read out. We again used the difference between two experiments,
where the two RF pulses have a phase difference of $\phi_{d}=\nu_{det}\tau_{f}$
and $\pi+\phi_{d}$ , respectively, to suppress unwanted background
signals. Fig.~\ref{fid_all} (b), shows the FIDs measured with a
detuning frequency of $\nu_{det}$=40 MHz, together with a fit to
the a function

\begin{equation}
S_{x+\phi_{d}}-S_{-x+\phi_{d}}=A\,cos(2\pi\nu_{det}\tau_{f}+\phi)e^{-\tau_{f}/T_{2}^{\text{*}}},\label{eq_fid}
\end{equation}
where $S_{x+\phi_{d}}(\tau_{f})$ and $S_{-x+\phi_{d}}(\tau_{f})$
are the average PL signals measured with the $\pm x+\phi_{d}$ detection
pulse. The fit parameter $T_{2}^{*}$=38 ns for V$_{1}$/V$_{3}$
and $T_{2}^{*}$=31 ns for V$_{2}$ at room temperature and in the
absence of an external magnetic field.

\subsection{Spin-echo}

The decay of the coherence is due to different types of interactions
that are broadly classified as homogeneous vs. inhomogeneous. They
can be separated by the spin-echo experiment (Hahn-echo)~\cite{2028}.
We will refer to the homogenous decay time of the Hahn echo as $T_{2}$.

\begin{figure}[h]
\includegraphics{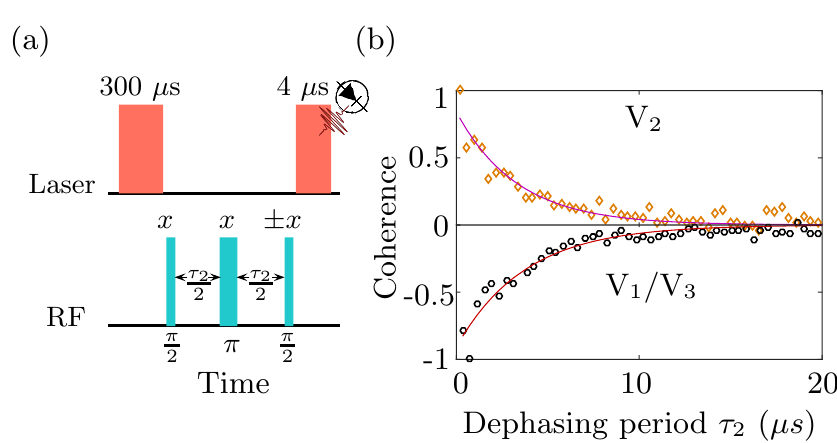}

\caption{(a) Pulse sequence used to measure dephasing of the transverse spin
components. The red and blue rectangles represent the laser- and RF
pulses, respectively. The length of the pulses is written above them.
(b) Signals measured for V$_{1}$/V$_{3}$ and V$_{2}$ as a function
of the delay $\tau_2$.}
\label{t2_all}
\end{figure}

Fig. \ref{t2_all} (a) shows the pulse sequence for measuring the
spin-echo relaxation. After the initializing laser-pulse, we applied
a $\pi/2$ RF pulse, which created the coherence that evolved freely
for a time $\tau_2/2$. We then applied a refocusing $\pi$ pulse,
and a second $\tau_2/2$ delay. The remaining coherence was converted
into population by the third RF pulse and read out during the final
laser pulse. In this sequence, all RF pulses were applied along the
$x$-axis. In the reference signal, the last $\pi/2$ pulse was applied
along the -$x$-axis, which changed the sign of the resulting population
difference. Subtracting the signals from the two experiments thus
yielded a background-free measurement of the coherence. Fig.\ref{t2_all}
(b), plots the resulting data as a function of the dephasing period
$\tau_{2}$, together with a fit to an exponential decay
\[
S_{x}-S_{-x}=A~e^{-\tau_{2}/T_{2}},
\]
where $S_{x}(\tau_2)$ and $S_{-x}(\tau_2)$ are the signals measured
with the $\pm x$ pulse. The resulting values for $T_{2}$ were $3.73\pm0.13$
$\mu$s and $3.31\pm0.24$ $\mu$s for the V$_{1}$/V$_{3}$ and V$_{2}$
centers at room temperature and in the absence of an external magnetic
field.

\subsection{Echo trains}

\begin{figure}[h]
\includegraphics{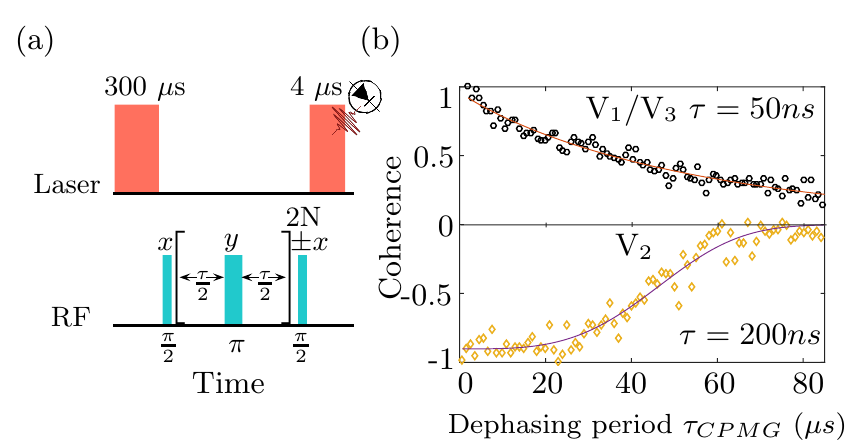}\caption{(a) Pulse sequence for measuring the spin coherence time under multiple
refocusing pulses. The red and blues rectangles represent the laser
and RF pulses respectively. (b) Decay of the spin coherence during
the multiple echo sequence. The experimental data (circles) are fitted
to function \eqref{eq:CPMGDecay}.}

\label{cpmg_all}
\end{figure}

Multiple refocusing pulses can extend the lifetime of the coherence
further, compared to the case of a single echo, if the perturbation
is not static, but its correlation time is longer than the spacing
between the echo pulses~\cite{carr-pr-1954,meiboom:688,RevModPhys.88.041001}.
We therefore measured the coherence time of the centers during a CPMG
pulse sequence, as shown in Fig.~\ref{cpmg_all} (a). After the initializing
laser-pulse, the $\pi/2$ RF pulse created spin coherence. During
the subsequent delay, we applied $2N$ refocusing pulses, each with
flip angle $\pi$. The final $\pi/2$ pulse converted the coherence
back into population which was read out as discussed above. To eliminate
background signals, we subtracted the results of the two experiments
where the final $\pi/2$ RF pulse rotated the spins around the $\pm x$
axes. The difference between the experimentally observed signals was
fitted to the function
\begin{equation}
S_{x}-S_{-x}=A\,e^{-\left(\tau_{CPMG}/T_{2}^{CPMG}\right)^{n}},\label{eq:CPMGDecay}
\end{equation}
where the total evolution period is $\tau_{CPMG}$=$2N\tau+2N$$\tau_{\pi pulse}$,
$2N$ is number of $\pi$ pulses and $\tau_{\pi pulse}$ is the duration
of a $\pi$ pulse, which was $17.5$ ns for V\textsubscript{1}/ V$_{3}$
and $21$ ns for V$_{2}$. Fig.~\ref{cpmg_all} (b) shows the experimental
data for a pulse spacing of $\tau=\,$$50$ ns for V\textsubscript{1}/V$_{3}$
and $\tau=$$200$ ns for V\textsubscript{2} . The fitted parameters
are $T_{2}^{CPMG}$ = $56\pm11$ $\text{ \ensuremath{\mu}s}$, \textit{$n=0.93$}
for V$_{1}$/V$_{3}$ and $T_{2}^{CPMG}$ = $51\pm4$ $\text{ \ensuremath{\mu}s}$,\textit{
$n\:$}= 3.47 for V$_{2}$ at room temperature and in the absence
of an external magnetic field.

\section{Discussion and Conclusion}

\label{conc}

V$_{Si}^{-}$ centers in SiC have interesting properties that may
become useful in future quantum devices, similar to the NV center
in diamond. In this work, we have studied in detail their properties
in the 6H-SiC polytype. We studied the photoluminescence spectrum
as a function of temperature. At low temperature, the ZPLs of the
V\textsubscript{1}, V\textsubscript{2} and V\textsubscript{3} are
quite sharp but broaden with increasing temperature.

These V$_{Si}^{-}$ centers can be spin-polarised by optical irradiation,
the spin can be manipulated by RF fields and read out optically .
In zero field, the ODMR spectrum shows two peaks with opposite amplitudes,
one at 28 MHz and the other at 128 MHz. They can be assigned to the
V$_{1}$/ V$_{3}$ and V$_{2}$ type V$_{Si}^{-}$~\cite{sorman-prb-00}.
ODMR transitions as a function of the magnetic field parallel to $c$-axis
were measured for both V$_{Si}^{-}$. 

The main emphasis of this work was on the coherence properties of
the spins during free precession as well as during the application
of refocusing sequences designed to protect the spins against environmental
noise. Using a train of echo pulses, we could extend the coherence
time of the V$_{Si}^{-}$ in the 6H-SiC polytype at room temperature
up to 50 $\mu$s.  Previous experimental room temperature studies
on V$_{2}$ type vacancy in 4H-SiC in the absence of external magnetic
field reported free evolution time and spin-echo decay times of 190-300
ns and 6-8 us respectively~\cite{carter-prb-15,simin-prb-17}. The
isotopic composition of the 4H-SiC samples used in those studies was
natural abundance, i.e. with 4.7 \% of $^{29}$Si and 1.1 \% of $^{13}$C,
both of which have nuclear spins $I=1/2$. Since the atoms closest
to the Si vacancy are all $^{13}$C nuclei, their hyperfine interaction
is approximately an order of magnitude stronger than that of the $^{29}$Si
nuclei located in the next-nearest neighbor (NNN) shell consisting
of twelve silicon atoms~\cite{wagner-prb-02,yang-prb-14,witzel-prb-05}.
The decoherence of the vacancy-spin ensemble is mainly due to dipolar
fluctuations of the nuclear spins~\cite{witzel-prb-05,yang-prb-14}
which couple to the electron spin couples by hyperfine interaction~\cite{yang-prb-14}.
A higher percentage of $^{13}$C therefore reduces the coherence time.
We expect that the coherence times can be extended by applying suitable
magnetic fields and dynamical decoupling schemes~\cite{carter-prb-15,simin-prb-17}.
Lowering the temperature will also extend the spin-lattice relaxation
time~\cite{simin-prb-17}.  These possibilities will be explored
in upcoming work.
\begin{acknowledgments}
This work was supported by the Deutsche Forschungsgemeinschaft in
the frame of the ICRC TRR 160 (Project No. C7)and by RFBR, project
number 19-52-12058. SIMS measurements were performed using the Center
of Multi-User Equipment ``Material Science and Diagnostics for Advanced
Technologies\char`\"{} (Ioffe Institute, Russia) facility supported
by the Russian Ministry of Science (The Agreement ID RFMEFI62119X0021).
\end{acknowledgments}

\end{document}